\documentclass[12pt]{iopart}
\pdfoutput=1

\usepackage{iopams}
\usepackage{graphicx}
\usepackage{bm}
\usepackage{cite}
\usepackage{color}
\usepackage[super]{nth}
\usepackage{booktabs}
\usepackage[squaren]{SIunits}
\usepackage{hyperref}
\hypersetup{colorlinks=true,citecolor=blue,linkcolor=blue,urlcolor=blue,pdfauthor={Ling Zhong, Edwin P Menzel, Roberto Di Candia, Peter Eder, Matthias Ihmig, Alexander Baust, Max Haeberlein, Elisabeth Hoffmann, Kunihiro Inomata, Tsuyoshi Yamamoto, Yasunobu Nakamura, Enrique Solano, Frank Deppe, Achim Marx, Rudolf Gross},pdftitle={Squeezing with a flux-driven Josephson parametric amplifier}}
\usepackage[all]{hypcap}
 
\begin{document}
\title[]{Squeezing with a flux-driven Josephson parametric amplifier}
\author{L~Zhong$^{1,2,9}$, E~P~Menzel$^{1,2,9}$, R~Di~Candia$^{3}$, P~Eder$^{1,2}$, M~Ihmig$^{5}$, A~Baust$^{1,2}$, M~Haeberlein$^{1,2}$, E~Hoffmann$^{1,2}$, K~Inomata$^{6}$, T~Yamamoto$^{6,7}$, Y~Nakamura$^{6,8}$, E~Solano$^{3,4}$, F~Deppe$^{1,2}$, A~Marx$^{1}$, and R~Gross$^{1,2}$}
\address{$^1$ Walther-Mei{\ss}ner-Institut, Bayerische Akademie der Wissenschaften, D-85748 Garching, Germany}
\address{$^2$ Physik-Department, Technische Universit\"{a}t M\"{u}nchen, D-85748 Garching, Germany}
\address{$^3$  Department of Physical Chemistry, University of the Basque Country UPV/EHU, Apartado 644, E-48080, Bilbao, Spain}
\address{$^4$ IKERBASQUE, Basque Foundation for Science, 48011 Bilbao, Spain}
\address{$^5$ Lehrstuhl f{\"u}r Integrierte Systeme, Technische Universit{\"a}t M{\"u}nchen, D-80333 M{\"u}nchen, Germany}
\address{$^6$RIKEN Center for Emergent Matter Science (CEMS), 2-1 Hirosawa, Wako, Saitama 351-0198, Japan}
\address{$^7$ NEC Smart Energy Research Laboratories, Tsukuba, Ibaraki, 305-8501, Japan}
\address{$^8$ Research Center for Advanced Science and Technology (RCAST), The University of Tokyo, Komaba, Meguro-ku, Tokyo 153-8904, Japan}
\footnote[0]{$^9$ These authors contributed equally to this work.}
\eads{\mailto{ling.zhong@wmi.badw.de}, \mailto{rudolf.gross@wmi.badw.de}}

\begin{abstract} 
Josephson parametric amplifiers (JPA) are promising devices for applications in circuit quantum electrodynamics (QED) and for studies on propagating quantum microwaves because of their good noise performance. In this work, we present a systematic characterization of a flux-driven JPA at millikelvin temperatures. In particular, we study in detail its squeezing properties by two different detection techniques. With the homodyne setup, we observe squeezing of vacuum fluctuations by superposing signal and idler bands. For a quantitative analysis we apply dual-path cross-correlation techniques to reconstruct the Wigner functions of various squeezed vacuum and thermal states. At $10\,\deci\bel$ signal gain, we find $4.9\pm0.2\,\deci\bel$ squeezing below vacuum. In addition, we discuss the physics behind squeezed coherent microwave fields. Finally, we analyze the JPA noise temperature in the degenerate mode and find a value smaller than the standard quantum limit for phase-insensitive amplifiers. 
\end{abstract}

\maketitle

\section{Introduction}
\label{section_introduction}

The tremendous progress in the field of quantum electrodynamics (QED) with solid-state superconducting circuits~\cite{Blais:2004, Wallraff:2004,Blais:2007,Schoelkopf:2008} has recently triggered massive efforts aiming at the investigation of propagating quantum microwaves~\cite{Mallet:2011,Menzel:2010,PhysRevLett.109.250502,Eichler:2011b,PhysRevLett.106.220503,PhysRevLett.109.240501,bozyigit2010antibunching,tien1958parametric,PhysRevLett.109.183901,Wilson:2011a}. Towards this end, the analysis of propagating microwave light at frequencies of a few gigahertz and with power levels below a single photon on average has become an important task. However, due to the low signal energy of typically only a few attowatt per megahertz bandwidth, the measurement of such signals requires amplification. For a long time, phase-insensitive high electron mobility transistor (HEMT) amplifiers are considered as a good choice. They feature a broad operation bandwidth, high gain, but still add 10-20 noise photons~\cite{Caves:1982,RevModPhys.82.1155} to the signal. This noise can be significantly reduced by using Josephson parametric amplifiers (JPA)~\cite{Yurke:1987,Yurke:1989a,Yurke:1996,Siddiqi:2004,Tholen:2007,Castellanos-Beltran:2007,Castellanos-Beltran:2008,Yamamoto:2008, Abdo:2009,eichler2013controlling}, which achieve amplification via a high-frequency modulation of a Josephson inductance. In the phase-insensitive or non-degenerate operation mode, the JPA noise temperatures approach the standard quantum limit dictated by the Heisenberg uncertainty relation~\cite{Movshovich:1990,Mallet:2011,Eichler:2011b,Siddiqi:2004,eichler2013controlling}. Even better, JPAs can in principle amplify a single signal quadrature without adding any noise~\cite{Caves:1982}. This property is intimately connected to the fact that a JPA can squeeze a single quadrature below the level of the vacuum fluctuations. Hence, in combination with lately developed state reconstruction methods for propagating quantum microwaves ~\cite{Menzel:2010,PhysRevLett.109.250502,PhysRevLett.106.220503}, a JPA is the ideal tool to study the rich physics of squeezed microwave states. Recently, the squeezed vacuum generated by the JPA investigated in this work was successfully used to generate distributable path entanglement~\cite{PhysRevLett.109.250502}. In addition, squeezing is considered as a key ingredient for the creation of quantum memories for entangled continuous-variable states\cite{Jensen:2011}. 

Here, we present a comprehensive study on the squeezing physics produced by a flux-driven JPA~\cite{Yamamoto:2008}. The particular advantage of this design is the large isolation between the pump and signal ports.  For this reason, the elimination of the pump tone becomes obsolete~\cite{Castellanos-Beltran:2007} and, thus, experimental complexity is greatly reduced. After characterizing the basic properties of our JPA with a spectroscopy setup and a standard homodyne detection scheme, we reconstruct the Wigner functions of the JPA output states using dual-path cross-correlation techniques~\cite{Menzel:2010}. The latter allows us to investigate squeezed vacuum, squeezed thermal and squeezed coherent states. We find a maximum squeezing of $4.9\pm0.2\,\deci\bel$ and confirm that the dependence of photon number and displacement on the squeezed and coherent state angle follow the expectations from theory. In addition, a detailed noise analysis shows that our JPA is operated near the quantum limit in the phase-insensitive mode and that its noise temperature is below the standard quantum limit in the degenerate mode. All in all, our work illuminates in unprecedented detail the fundamental physics of single-mode squeezing in the microwave domain.

The paper is organized as follows: In section~\ref{section_JPA_characterization}, we discuss the characterization of the flux-driven JPA using spectroscopy and homodyne detection. After introducing sample and measurement setups, we analyze signal and idler gain, the associated bandwidths, compression effects, and the noise properties in non-degenerate mode. Next, we investigate in section~\ref{section_squeezing} the squeezing of vacuum and thermal states with both homodyne detection and the dual-path state reconstruction method. In section~\ref{section_squeezed_coherence}, we extend our discussion to squeezed coherent states by operating the JPA in the degenerate mode with coherent input signals. After discussing the JPA noise properties in the degenerate mode based on dual-path measurements in section~\ref{section_noise}, we summarize our main results and give a short outlook in section~\ref{section_conslusions}.

\section{JPA characterization}
\label{section_JPA_characterization}

\subsection{JPA sample}
\label{subsection_Sample}

\begin{figure}[t!]
\centering{\includegraphics[width=1\columnwidth]{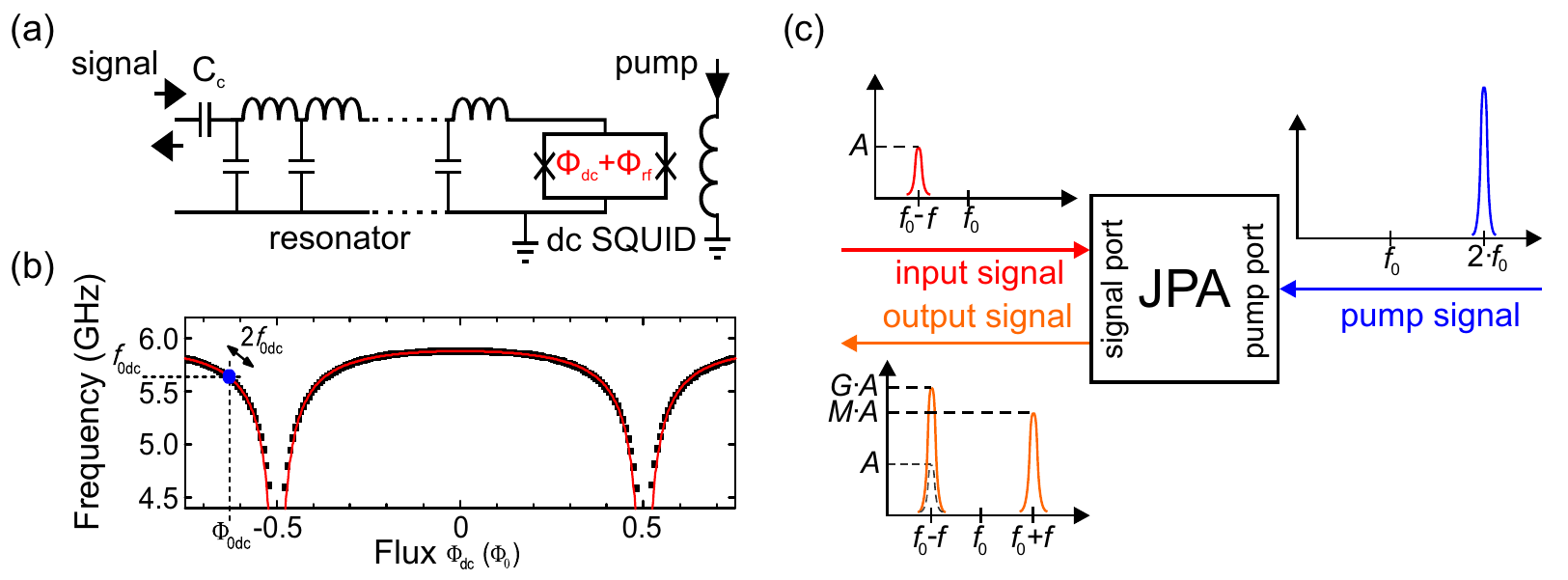}} 
\caption{Flux driven JPA. 
(a) Circuit diagram. The transmission line resonator is terminated by a dc SQUID (loop with crosses symbolizing Josephson junctions) at one end. A magnetic flux $\Phi_{\textrm{dc}}+\Phi_{\textrm{rf}}$ penetrating the dc SQUID modulates the resonant frequency. 
(b) Dependence of the resonant frequency on the dc flux. The red line is a fit of a distributed circuit model~\cite{PhysRevB.74.224506} to the data~(\fullsquare). Blue dot: Operation point used in our experiments. (c) Schematic of the operating principle of the JPA (see text for details).}
\label{fig:operation principle}
\end{figure}

In this subsection, we describe the operation principle of a flux-driven JPA in general and our sample in particular. A parametric amplifier is an oscillator whose resonant frequency is modulated periodically in time. In the case of the JPA, the oscillating system is a quarter-wavelength transmission line resonator whose resonant frequency  is determined by its capacitance and inductance (see figure~\ref{fig:operation principle}(a)). The latter can be varied by a dc superconducting quantum interference device (SQUID), which consists of a superconducting loop interrupted by two Josephson junctions and acts as a flux-dependent non-linear inductor. Thus, by modifying the magnetic flux $\Phi_{\rm dc}$ threading the SQUID loop, the resonant frequency can be adjusted (see figure~\ref{fig:operation principle}(b)). By fitting a physical model to the experimental data (black squares), we can estimate a Josephson coupling energy $E_{\rm J}\,{=}\,h\,{\times}\,1305\,\giga\hertz$ for each junction, where $h$ is the Planck constant. The slightly different values compared to those in the supplementary material of~\cite{PhysRevLett.109.250502} result from the fact that, here, we employ a more sophisticated distributed-element model~\cite{PhysRevB.74.224506} instead of a simple lumped-element approach.

Periodically varying the resonant frequency with an ac flux (pump tone) at $2f_0$, where $f_0$ is the operation point frequency, results in parametric amplification: A signal at $f_0-f$ impinging at the signal port is amplified by the signal gain $G$ and reflected back out of the signal port. At the same time, an idler mode at $f_0+f$ is created, whose amplitude is determined by the intermodulation gain $M$. This operation principle is depicted in figure~\ref{fig:operation principle}(c). If the incoming signal consists of vacuum fluctuations, this process is the analogue of parametric downconversion in optics, where a pump photon is split into a signal and an idler photon. Therefore, strong quantum correlations between the signal and idler mode are established which finally result in squeezing. 

\begin{figure}[t!]
\centering{\includegraphics[width=0.6\columnwidth]{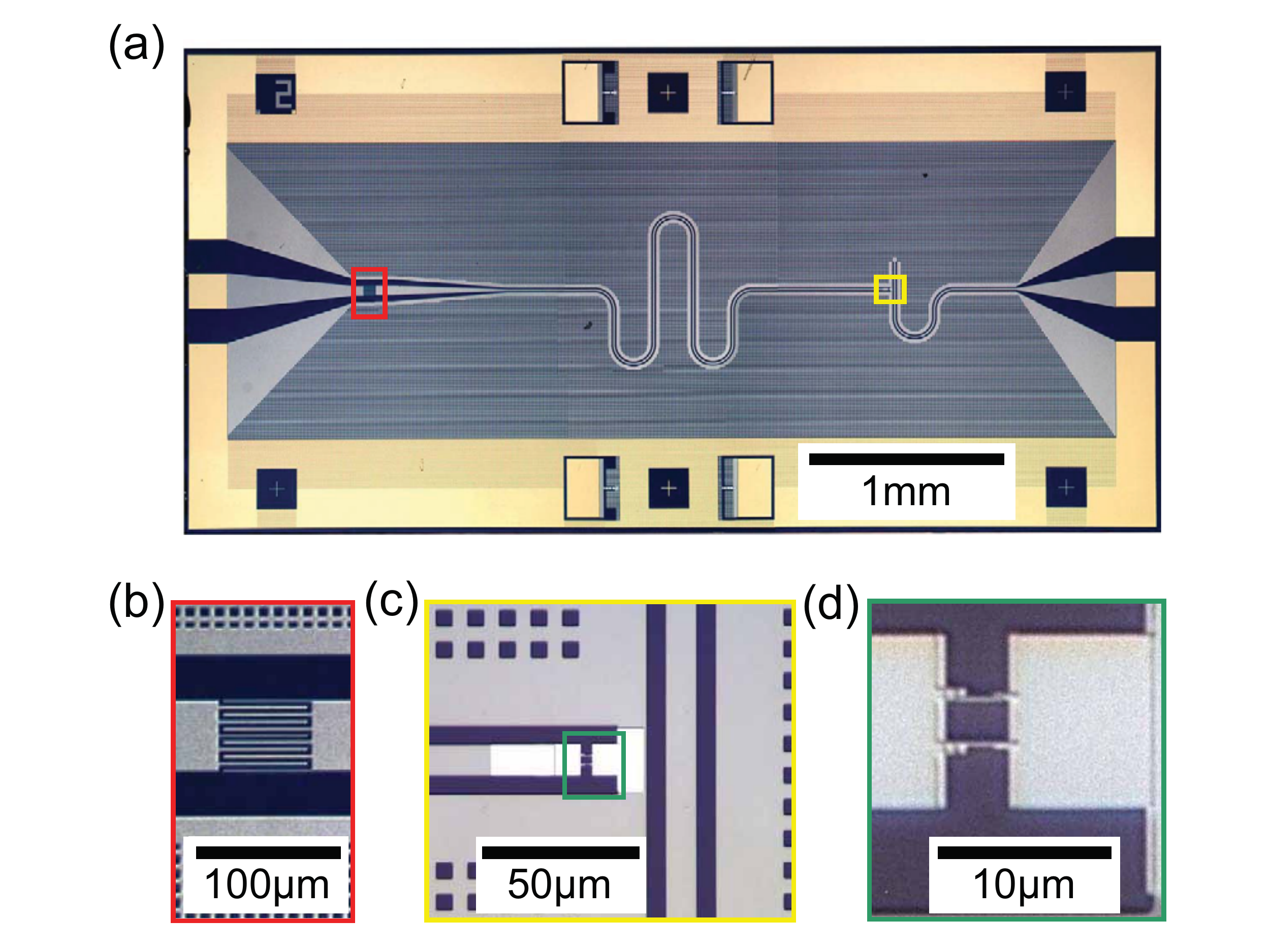}}
\caption{Flux driven JPA used in our experiments.
(a)~Optical micrograph of the chip. Red rectangle: coupling capacitor with the design coupling capacitance of $30\,\femto\farad$. Yellow rectangle: dc SQUID and pump line. 
(b)~Zoom-in of the coupling capacitor marked with the red rectangle in panel (a).
(c)~dc SQUID and pump line in the region marked with the yellow rectangle in panel (a).
(d)~Zoom-in of the dc SQUID marked with the green rectangle in panel (c). The size of the SQUID loop is $4.2\times2.4$\,$\mathrm{\micro\meter}^2$. 
}
\label{fig:sample}
\end{figure}

An optical micrograph of the sample is shown in figure~\ref{fig:sample}. The resonator and antenna are patterned from a sputtered $50\,\nano\meter$ thick Nb film. At the contacts, $95\,\nano\meter$ of gold is deposited on top of a $5\,\nano\meter$ titanium bonding layer. As substrate we use thermally oxidized ($300\,\nano\meter$) silicon with a thickness of $300\,\micro\meter$. In the last step, the aluminum dc SQUID is fabricated using shadow evaporation~\cite{dolan1977offset}. Each Al electrode has a thickness of $50\,\nano\meter$. The sample chip is placed between two alumina printed circuit boards inside a gold-plated copper box.

\subsection{Spectroscopy and homodyne setup}
\label{subsection_Setup}

\begin{figure}[t!]
\centering{\includegraphics[width=0.8\columnwidth]{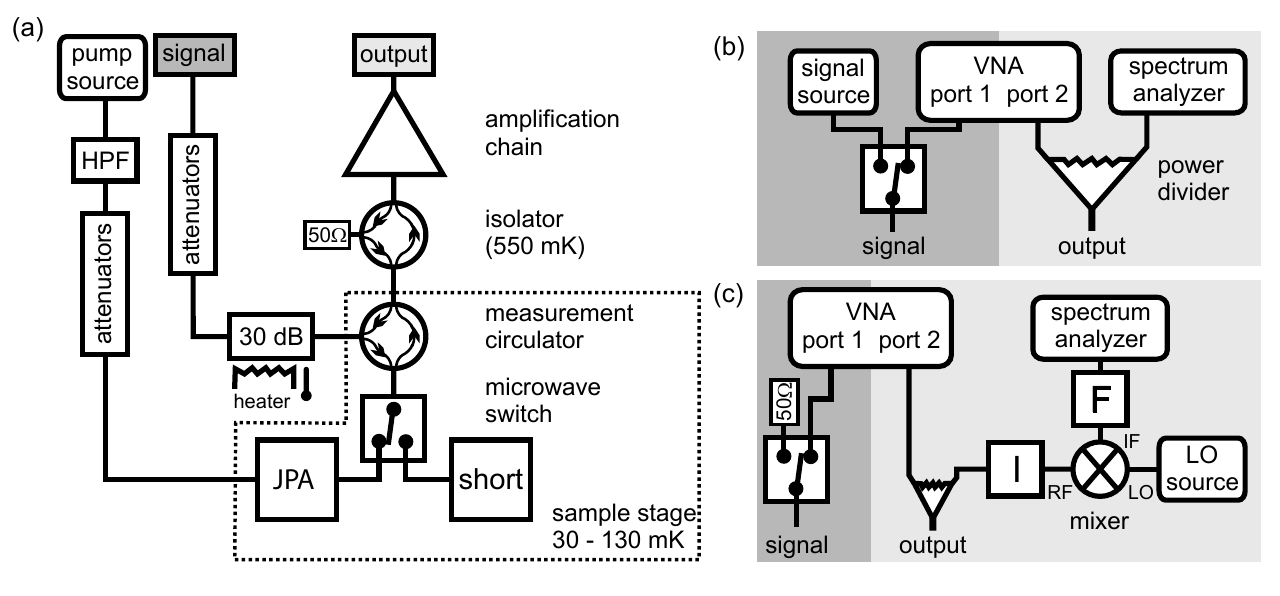}} 
\caption{
Schematics of the measurement setups for the JPA characterization
(a)~Cryogenic setup. The signal and pump line attenuators and the amplification chain partially extend to room temperature. HPF stands for high pass filter.
(b)~Room temperature spectroscopy setup for the characterization of the amplification and noise properties. 
(c)~Room temperature homodyne setup for the squeezing measurements. F~indicates a filter and I~an isolator. RF, IF and LO denote radio frequency, intermediate frequency and local oscillator, respectively.
}
 \label{fig:setup}
\end{figure}

The spectroscopy setup used to characterize the JPA and the homodyne setup to detect squeezing are sketched in figure~\ref{fig:setup}. In JPA characterization and homodyne measurements, the same cryogenic setup (see figure~\ref{fig:setup}(a)) is used. The sample is anchored to the sample stage of a dilution refrigerator and its temperature ranges between $90\,\milli\kelvin$ and $130\,\milli\kelvin$. The signal, generated either by a microwave source or a vector network analyzer (VNA), passes a series of warm (-40 to -$60\,\deci\bel$) and cold attenuators ($-69\,\deci\bel$) as shown in figure~\ref{fig:setup}(a). The signal power levels stated in this work are referred to the output of the 30\,\deci\bel-attenuator in figure~\ref{fig:setup}(a), while the pump power level is estimated at the input of the JPA sample box. The transmission from the $30\,\deci\bel$-attenuator output to the spectrum analyzer has been determined by sweeping the temperature of the $30\,\deci\bel$-attenuator and measuring the power of the emitted black body microwave radiation~\cite{PhysRevLett.105.133601}. We subtract this value from the total transmission between the microwave source output and the spectrum analyzer to calibrate for the loss of the input line. For signal and idler gain measurements, a coherent signal is fed through the attenuated input line via the measurement circulator to the JPA. This circulator separates the outgoing from the incoming signal and protects the JPA from the noise generated by the amplification chain. The amplified output signal can be detected by a VNA or a spectrum analyzer (see figure~\ref{fig:setup}(b)). The VNA allows to investigate the complex reflection coefficient of the JPA. The spectrum analyzer is used to investigate the idler gain and the degenerate operation of the JPA (see subsection~\ref{sec:degenerate gain}). Figure~\ref{fig:setup}(c) shows the homodyne receiver used in the squeezing measurements presented in section~\ref{section_squeezing}. It consists of a local oscillator microwave source and a mixer which downconverts the signal and idler modes onto the same intermediate frequency (IF). The resulting interference reveals the strong correlations established between the idler and signal modes. If the interference is destructive, fluctuations can be squeezed below the level of the vacuum fluctuations.

\subsection{Operation Point}
\label{Operation Point}

\begin{figure}[t!]
\centering{\includegraphics[width=0.7\columnwidth]{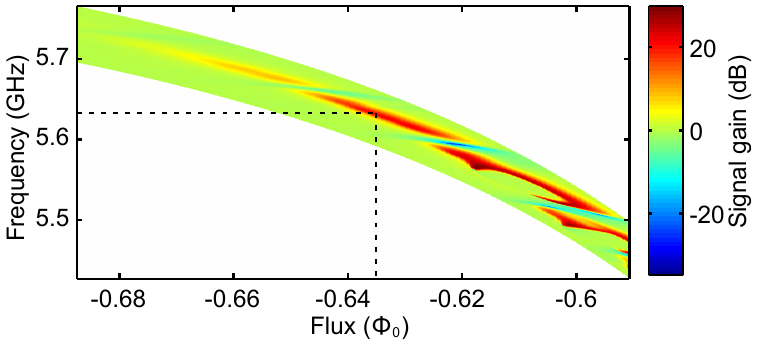}} 
\caption{
Signal gain as a function of frequency and applied magnetic flux at a pump power of $P_{\rm pump}\,{=}\,{-}39\,\deci\bel\milli$. The temperature of the JPA is stabilized at $130\,\milli\kelvin$. The dashed lines indicate the working point for our experiments. 
}
 \label{fig:operation point}
\end{figure}

The first step in characterizing the JPA is to determine a suitable operating point. To this end, we measure the signal gain with a VNA while synchronously sweeping the pump tone, fulfilling the relation $f_{\rm pump}\,{=}\,2f_{\rm signal}\,{+}\,10\,\kilo\hertz$. Since the VNA measurement bandwidth of $30\,\hertz$ is much smaller than $10\,\kilo\hertz$, only the signal mode is detected preventing interference effects present in the degenerate mode. We emphasize that this measurement is different from the signal bandwidth measurements discussed in subsection~\ref{subsection_the bandwidth}, where the pump tone is at a fixed frequency and only the signal frequency is swept. In figure~\ref{fig:operation point}, we have plotted the measured signal gain when synchronously sweeping the pump and signal frequency for different flux values. For lower frequencies, the signal gain is increasing because the dependence of the resonant frequency on the flux becomes steeper (see also figure~\ref{fig:operation principle}(b)). We choose our operating point $f_0$ between $5.634\,\giga\hertz$ and $5.639\,\giga\hertz$, depending on the measurement. As it can be seen from figure~\ref{fig:operation point}, our operation point is located in the center of a region where the signal gain is appreciable and its frequency dependence is well behaved. At this operation frequency, the external quality factor is measured as $Q_{\rm ext}\,{=}\,312$, and the isolation between antenna and resonator is at least $28\,\deci\bel$.

\subsection{Non-degenerate gain}
\label{subsection_the nondegenerate gain}

\begin{figure}[t!]
\centering{\includegraphics[width=0.7\columnwidth]{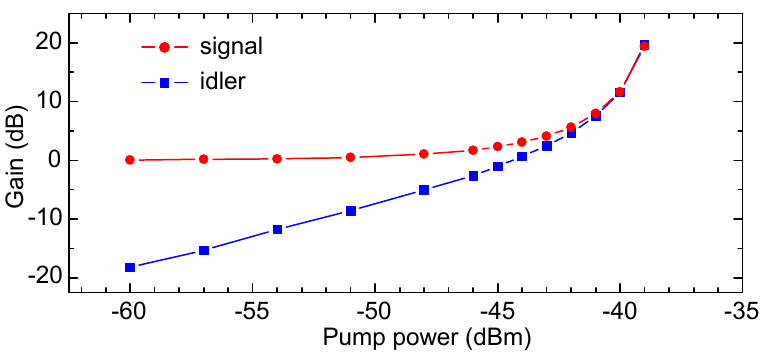}} 
\caption{
Pump power dependence of the idler and signal gain in the non-degenerate operation mode. The solid lines are guides to the eye. The temperature of the JPA is stabilized at $90\,\milli\kelvin$.
}
 \label{fig:gain pump power}
\end{figure}

When the signal frequency is detuned from half the pump frequency, signal and idler modes are at different frequencies and can be observed individually. This mode of operation is therefore called ``non-degenerate mode''. Figure~\ref{fig:gain pump power} shows the pump power dependence of the non-degenerate signal and idler gain for a detuning of $10\,\kilo\hertz$. For low pump power, no significant signal gain is observed and the idler gain is small. For large pump power, the two curves converge and both idler and signal gain reach $19.5\,\deci\bel$. 

\subsection{Bandwidth}
\label{subsection_the bandwidth}

\begin{figure}[t!]
\centering{\includegraphics[width=0.9\columnwidth]{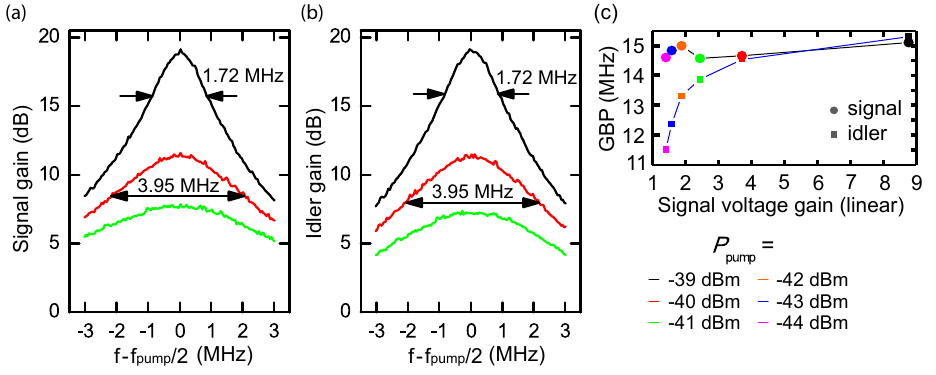}} 
\caption{
Signal (a) and idler (b) gain as a function of frequency, showing the bandwidth for different values of the pump power. (c)~GBP as a function of the signal voltage gain in linear units. The dashed lines are guides to the eyes. The temperature of the JPA is stabilized at $90\,\milli\kelvin$. 
}
 \label{fig:bandwidth pump}
\end{figure}

Besides the gain properties described above, bandwidth is an important feature of an amplifier. Therefore, we determine the instantaneous bandwidth of signal and idler modes at a fixed operation point (constant flux and pump frequency) by measuring the signal and idler gain for various detuning between half the pump frequency $f_{\rm pump}/2$ and the signal frequency $f$. As shown in figure~\ref{fig:bandwidth pump}, we observe a signal and idler bandwidth of $1.72\,\mega\hertz$ for a pump power of $-39\,\deci\bel\milli$. We define the gain-bandwidth-product (GBP), which is defined as the product of the voltage gain in linear units and the bandwidth of our amplifier. In the large gain limit, the GBP is nearly constant and close to the theoretical limit of $f_0/Q_{\rm ext}\,{=}\,18\,\mega\hertz$~\cite{Baust:2010} (see figure~\ref{fig:bandwidth pump}(c)). Going to low values of the signal gain, the idler gain vanishes and the signal gain approaches one because the signal gain is normalized to the pump-off condition. Therefore, we expect the idler GBP to decrease and signal GBP to increase. However, in the low gain limit the signal bandwidth measurement is very sensitive to the calibration data and fluctuations. Consequently, the signal GBP does not diverge in contrast to expectations from theory.

\subsection{\texorpdfstring{$1\,\deci\bel$}{1 dB}-compression point}
\label{subsection_compression point}

\begin{figure}[ht]
\centering{\includegraphics[width=0.75\columnwidth]{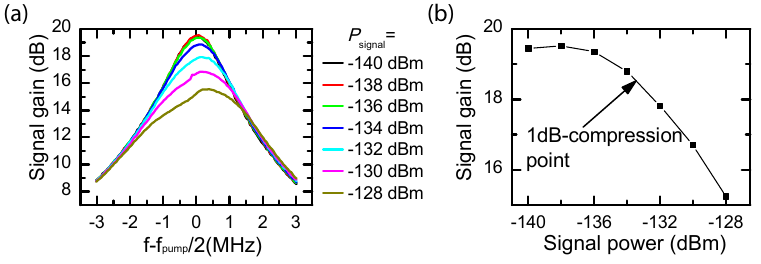}} 
\caption{
(a) Signal gain as a function of frequency for different values of the signal power and (b)~signal gain at $f_{\rm pump}/2\,{+}\,10\,\kilo\hertz$ versus signal power. Data: \fullsquare. Line: Guide to the eye. The temperature of the JPA is stabilized at $88\,\milli\kelvin$.
}
 \label{fig:compression}
\end{figure}

Another important figure of merit for amplifiers is their $1\,\deci\bel$-compression point. It denotes the power where the signal gain is $1\,\deci\bel$ below the value expected for a perfectly linear device~\cite{eichler2013controlling}. In other words, at some point the signal gain starts to decrease as a function of input power due to the nonlinearity of the amplifier. In figure~\ref{fig:compression}(a) bandwidth measurements of the signal gain are displayed. In contrast to figure~\ref{fig:bandwidth pump}, here the pump power is fixed at -39\,$\deci\bel\milli$ and the dependence on the signal power is studied. For small signal powers ($P_{\rm signal}\,\leq\,-136\,\deci\bel\milli$) the curves overlap meaning that the gain is constant and that the amplifier is in the linear regime. For larger signal powers a reduction of the gain is observed. Figure~\ref{fig:compression}(b) shows the maxima of the signal gain of figure~\ref{fig:compression}(a) for several signal powers. The $1\,\deci\bel$-compression point occurs at $-133\,\deci\bel\milli$. An analogous analysis for the idler gain (data not shown) reveals that the compression occurs for both the idler and signal gain at the same input power. For a circuit QED experiment with a cavity decay rate of $1\,\mega\hertz$, this power level is equivalent to 10 photons on average.

\subsection{JPA noise properties in non-degenerate mode}
\label{subsection_noise homodyne}

\begin{figure}[ht]
\centering{\includegraphics[width=0.7\columnwidth]{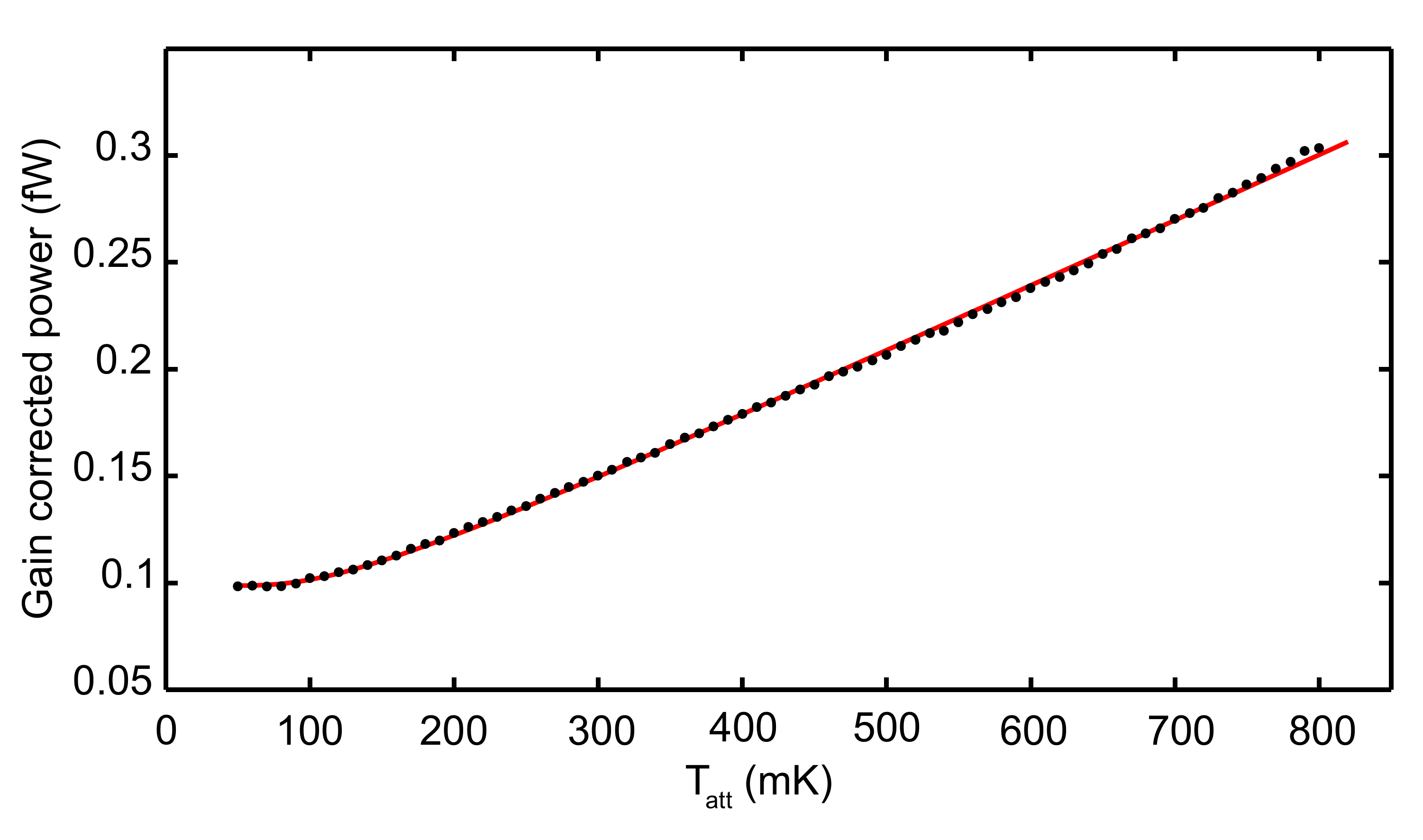}} 
\caption{
Gain corrected power as a function of the noise source temperature. Red line: Fit to the data (\fullcircle$\!\!$) using~(\ref{eq:BosePlanck}). The JPA temperature is in the range from $92$ to $115\,\milli\kelvin$.
}
\label{fig:noise temperature}
\end{figure}

A low noise temperature of an amplifier is very important when using it for measuring signals at the quantum level. Here, we use the spectroscopy setup for a rough characterization of the noise properties of our JPA when it is operated in the phase-insensitive mode. For details on the noise temperature in the degenerate mode, we refer to section~\ref{section_noise}. In order to determine the JPA noise properties, we measure the amplified noise power emitted by the $30\,\deci\bel$-attenuator, whose temperature is controlled in the range $T_{\rm att}\,{=}\,50\,{-}\,800\,\milli\kelvin$. The total noise power $P$ at the spectrum analyzer is given by
\begin{equation}
P(T_{\rm att})=G B\left\{\frac{h f_0}{2} \coth\left[\frac{h f_0}{2 k_{\rm B}\left( T_{\rm att}+\delta T\right)}\right] + k_{\rm B}T_{\rm total}\right\},
\label{eq:BosePlanck}
\end{equation}
where $G$ denotes the total gain, $B$ the detection bandwidth, $h=6.626\times 10^{-34}$\,J$\cdot$s the Planck constant, $k_{\rm B}\,{=}\,1.38\times 10^{-23}\,\joule/\kelvin$ the Boltzmann constant and $T_{\rm total}$ the total noise temperature of the complete detection chain, which includes the measurement circulator, the JPA and the amplification chain. $T_{\rm total}$ is related to the number of photons $n_{\rm total}$ added by the complete detection chain by $k_{\rm B} T_{\rm total}\,{=}\,n_\mathrm {total}h f_0$. The first term in (\ref{eq:BosePlanck}) describes thermal fluctuations and vacuum fluctuations according to~\cite{Callen:1951}. Possible deviations between the electronic temperature of the attenuator and the measured temperature are taken into account by $\delta T$. We set the signal gain to $G_{\rm signal}(T{\rightarrow}0)\,{=}\,19\,\deci\bel$ at the operation point $f_0$. However, sample heating due to the pumping process and compression effects at high noise source temperatures may cause the signal gain to deviate from this value. Therefore, we measure the dependence of the gain-corrected power on the noise source temperature $T_{\rm att}$ (see figure~\ref{fig:noise temperature}). To this end, we implement the following protocol for each temperature point: after measuring the signal gain with the VNA, we turn off the VNA and measure the total noise power using a spectrum analyzer. We obtain the gain corrected power by dividing the total noise power at each temperature point by the effective JPA gain for white uncorrelated noise, $G_{\rm{eff}}(T)\,{=}\,2G_{\rm{signal}}(T)\,{-}\,1$. Here, we consider the idler gain using $G_{\rm{idler}}\,{=}\,G_{\rm{signal}}\,{-}\,1$~\cite{Yurke:1989a}. Taking into account the cable and connector losses between the noise source and the measurement circulator, we estimate $T_\mathrm {total}\,{=}\,167\,\milli\kelvin$, which corresponds to $n_\mathrm {total}\,{=}\,0.62$ for the noise photons added by the whole detection chain referred to the input of the measurement circulator. This value is close to the standard quantum limit for phase-insensitive amplifiers of 0.5 photons ($135\,\milli\kelvin$). In addition, we find $\delta T\,{=}\,-19\,\milli\kelvin$ to be reasonably small. 

\subsection{Degenerate gain}
\label{subsection_the degenerate gain}

\begin{figure}[ht]
\centering{\includegraphics[width=0.7\columnwidth]{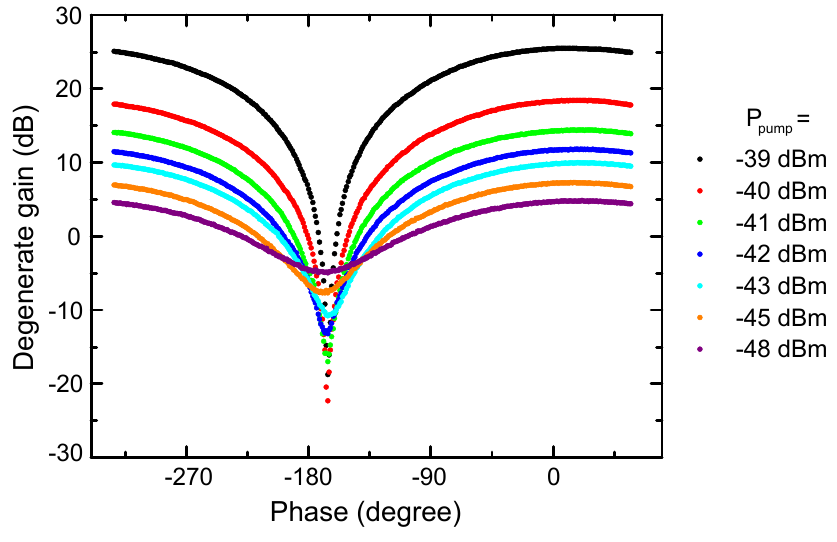}} 
\caption{
Degenerate gain as a function of the phase difference between probe and pump signal for different values of the pump power. For clarity, the curves are shifted in phase direction, so that the minima coincide. The temperature of the JPA is stabilized at $95\,\milli\kelvin$.
}
 \label{fig:degenerate gain}
\end{figure}

In order to demonstrate the usability of our JPA as a phase-dependent amplifier, we investigate the degenerate gain. In this mode of operation, the pump frequency is twice the signal frequency. Thus, the idler mode is created at the frequency of the amplified signal mode. This results in an interference of the two modes which is constructive or destructive depending on the phase between the idler and the signal modes, which can be controlled by shifting the phase difference between the probe signal and the pump tone. The degenerate gain is measured with a spectrum analyzer, where a measurement with zero pump power is used as a reference. 
In figure~\ref{fig:degenerate gain}, the degenerate gain is plotted as a function of the phase between the probe signal and the pump for different pump power levels. We observe a maximum degenerate gain of $25.5\,\deci\bel$ and a maximum deamplification of $22.3\,\deci\bel$. The former value is consistent with the signal and idler gain of $19.5\,\deci\bel$ (see Sec.~\ref{subsection_the nondegenerate gain}), since the constructive interference of equal amplitudes results in a $6\,\deci\bel$ increase of gain. 

\label{sec:degenerate gain}

\section{Squeezing of vacuum and thermal fluctuations}
\label{section_squeezing}

As mentioned in the discussion of the operation principle (subsection~\ref{subsection_Sample}), the JPA creates quantum correlations between the signal and idler modes. In the degenerate operation mode, these correlations result in deamplification or amplification depending on the quadrature direction. In the case of deamplification, the quadrature fluctuations can be squeezed below those of the vacuum. In this section, we investigate squeezed vacuum fluctuations generated by the flux-driven JPA with two different detection techniques: homodyne detection and dual-path state reconstruction~\cite{Menzel:2010,PhysRevLett.109.250502}.

To define the squeezing level $\cal S$ in decibel, we compare the variance of the squeezed quadrature $\left(\Delta X_{\rm sq}\right)^2$ with the quadrature variance of vacuum fluctuations, obtaining
$$
{\cal S}=\max\{0,-10 \lg\left[\left(\Delta X_{\rm sq}\right)^2/0.25\right]\}.
$$
We note that $\left(\Delta X_{\rm sq}\right)^2\,{<}\,0.25$ indicates squeezing and $\cal S$ is positive. Larger $\cal S$ corresponds to a higher squeezing level. $\left(\Delta X_{\rm sq}\right)^2\geq0.25$ indicates no squeezing and $\cal S$ equals zero. Hence, in this work we use the nomenclature that the term ``squeezing'' is equivalent to ``squeezing below the vacuum level''. 

\subsection{Squeezing detected with the homodyne setup }
\label{subsection_homodyne squeezing}

\begin{figure}[t!]
\centering{\includegraphics[width=0.7\columnwidth]{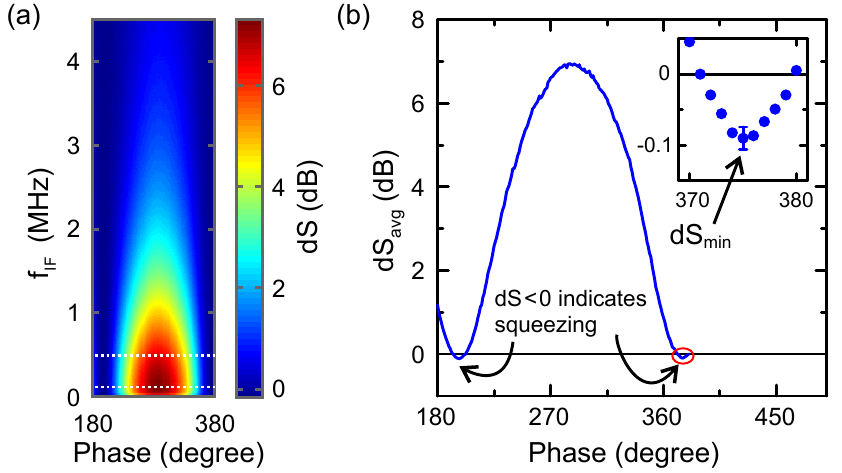}} 
\caption{
Squeezing of vacuum fluctuations detected with the homodyne setup. (a)~Power spectral density ratio $dS$ plotted as a function of phase and intermediate frequency. (b)~The power spectral density ratio averaged over the frequency range from 118 to $487\,\kilo\hertz$, $dS_{\rm avg}$, plotted as a function of phase. The averaged region is indicated by the dotted lines in panel~(a). The inset shows a zoom-in of the region marked by the red ellipse. The JPA temperature is stabilized at $88\,\milli\kelvin$.
}
\label{fig:squeezing}
\end{figure} 
First, we detect the output signal of the JPA with the homodyne detector shown in figure~\ref{fig:setup}(c). By sweeping the temperature of the $30\,\deci\bel$-attenuator from $50\,\milli\kelvin$ to $800\,\milli\kelvin$ with inactive JPA, we calibrate the gain and the noise temperature of the detection chain using (\ref{eq:BosePlanck}). Then, we cool the attenuator to $33\,\milli\kelvin$ and investigate the squeezed vacuum state. In figure~\ref{fig:squeezing}(a), the ratio ${dS}\equiv S_{\rm{on}}/S_{\rm{off}}$ is plotted as a function of the relative phase between the pump and the local oscillator and the detected intermediate frequency $f_{\rm IF}$. The latter is obtained by  downconverting the signal using a fixed local oscillator frequency $f_{\rm LO}=f_0=f_{\rm pump}/2$. For these settings, signals at $f_{\rm LO}\,{+}\,f_{\rm IF}$ and $f_{\rm LO}\,{-}\,f_{\rm IF}$, representing the signal and idler modes, are downconverted to the same intermediate frequency $f_{\rm IF}$. Therefore, the homodyne detector is sensitive to the correlations between the two modes created by the JPA. Here, $S_{\rm on}$ and $S_{\rm off}$ are the power spectral densities recorded with the JPA pump power on and off, respectively. In figure~\ref{fig:squeezing}(b), the average $dS_{\rm avg}$ of $dS$ calculated in the frequency range $118\,\kilo\hertz\,{<}\,f_{\rm IF}\,{<}\,487\,\kilo\hertz$ is plotted. Whenever $dS_{\rm avg}\,{<}\,0\,\deci\bel$, the noise detected at the spectrum analyzer referred to the input of the amplification chain is smaller than the vacuum noise emitted by the attenuator. This demonstrates the effect of vacuum squeezing. From the minimal value $ dS_{\rm min}$ indicated in the inset of figure~\ref{fig:squeezing}(b), we calculate the squeezing level ~\cite{Movshovich:1990}
$$
{\cal S}=-10\lg\left[1-\frac{T_{\rm n} \left(1-10^{dS_{\rm min}/10}\right)}{0.5 h f_0 /k_{\rm B}}\right].
$$
Here, $T_{\rm n}$ is the noise temperature of the detection chain with the JPA off. Taking into account the cable losses, reference spectrum fluctuation and thermal population at the input of the JPA, we retrieve a lower bound of $2.8\,\deci\bel$ of squeezing at the input of the amplification chain.

\subsection{Dual-path setup }
\label{subsection_dual path setup}

\begin{figure}[ht]
\centering{\includegraphics[width=0.6\columnwidth]{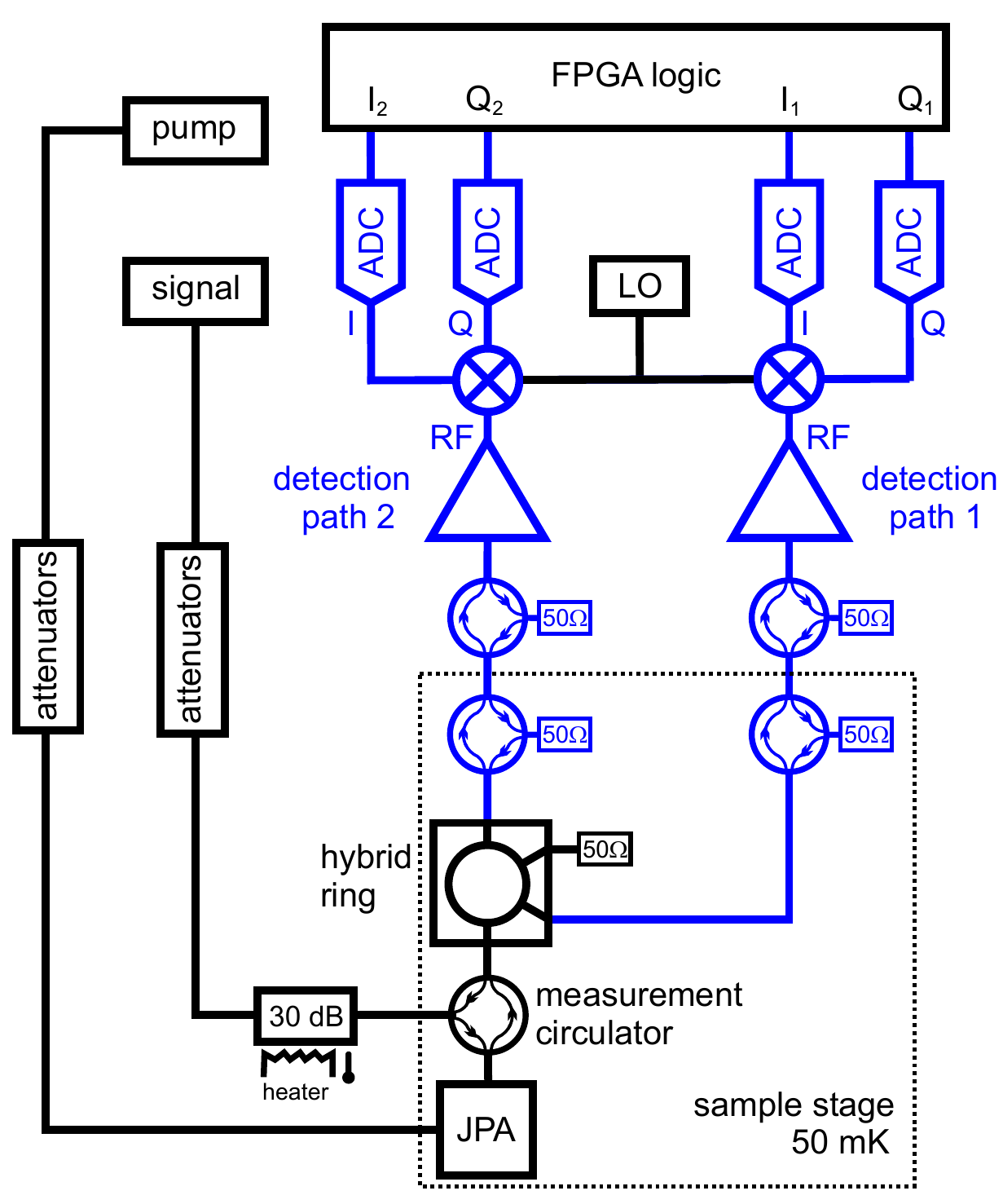}} 
\caption{
Simplified schematic of the dual path setup. The triangular symbols denote the amplification chains, the circles with crosses are IQ-mixers and the boxes labeled ADC are analog-to-digital converters. LO, pump, and signal denote the microwave sources for local oscillator, pump, and signal frequency. The heatable $30\,\deci\bel$-attenuator is thermally weakly coupled to the sample stage. In addition to the temperature control of this attenuator, the JPA sample temperature can be stabilized independently.
For simplicity, isolators, IF amplifiers and filters are omitted in the sketch.}
 \label{fig:DP setup}
\end{figure}

The estimation of the squeezing level with the homodyne detection setup only yields a lower bound for the amount of squeezing. Rather than improving this setup, we choose to fully reconstruct the squeezed state emitted by the JPA with the dual-path setup, which is based on cross-correlation techniques and realtime data processing~\cite{Menzel:2010,PhysRevLett.109.250502}. First, we introduce the dual-path setup in figure~\ref{fig:DP setup}. We split the signal under study described by the bosonic annihilation and creation operators $\hat{a}$ and $\hat{a}^\dagger$ using a microwave beam splitter and feed them into two amplification and detection paths. During the splitting process vacuum fluctuations are added to the split signals. While the signal is emitted from our JPA sample, the vacuum fluctuations are realized by terminating the other input port of the beam splitter with a broadband $50\,\Omega$ load. At the two output ports, we first amplify the signals using cold HEMT and room temperature amplifiers and then downconvert them to an intermediate frequency (IF) of $11\,\mega\hertz$ using IQ-mixers. The resulting orthogonal quadrature signals $I_{1,2}$ and $Q_{1,2}$ are then digitized by four analog-digital-converters (ADCs), and processed in real time by a field programmable gate array (FPGA) logic. Based on the beam splitter relations and the fact that the noise contributions from two detection paths are independent, we get access to all quadrature moments of signal and the two detection paths up to fourth order. In the setup, we still have the same temperature-controllable $30\,\deci\bel$-attenuator at the end of the JPA input line to calibrate the photon number conversion factors which relate the autocorrelations measured in our detector (in units of $V^2$) to photon numbers at the input of the beam splitter.

After post-processing the ensemble averages of the noisy quadrature moments, the reconstructed signal and noise moments have the form of $\left<\left(\hat{a}^\dagger\right)^l \hat{a}^m\right>$ and $\left<\hat{V}_{1,2}^r\left(\hat{V}_{1,2}^\dagger\right)^s\right>$, respectively. Here, $\hat{V}_{1,2}$ and $\hat{V}_{1,2}^\dagger$ represent annihilation and creation operators of the noise mode in the detection path 1 and path 2, and $l$, $m$, $r$, $s$ $\in \{0, 1, 2, 3, 4\}$, with $l+m\le 4$ and $r+s\le 4$. Following~\cite{PhysRevLett.109.250502}, with the third and fourth moments we calculate the third and fourth order cumulants to verify the Gaussianity of the state. Furthermore, we use the first two moments to reconstruct the signal Wigner functions at the input of the beam splitter. 
At the same time, the noise moments of the two detection paths are obtained. We find that the detection chains add $24.22\pm0.02$ and $27.32\pm0.03$ photons referred to the beam splitter input which corresponds noise temperatures of $6.55\pm0.01\,\kelvin$ and $7.39\pm0.01\,\kelvin$ for the detection path 1 and path 2, respectively. Again, the error bars we provide are of purely statistical nature. In all dual-path experiments, the temperature of the JPA sample is stabilized at $50\,\milli\kelvin$.

\subsection{Squeezing detected with the dual-path setup }
\label{subsection_dual path squeezing}
\begin{figure}[t!]
\centering{
\includegraphics[width=0.9\columnwidth]{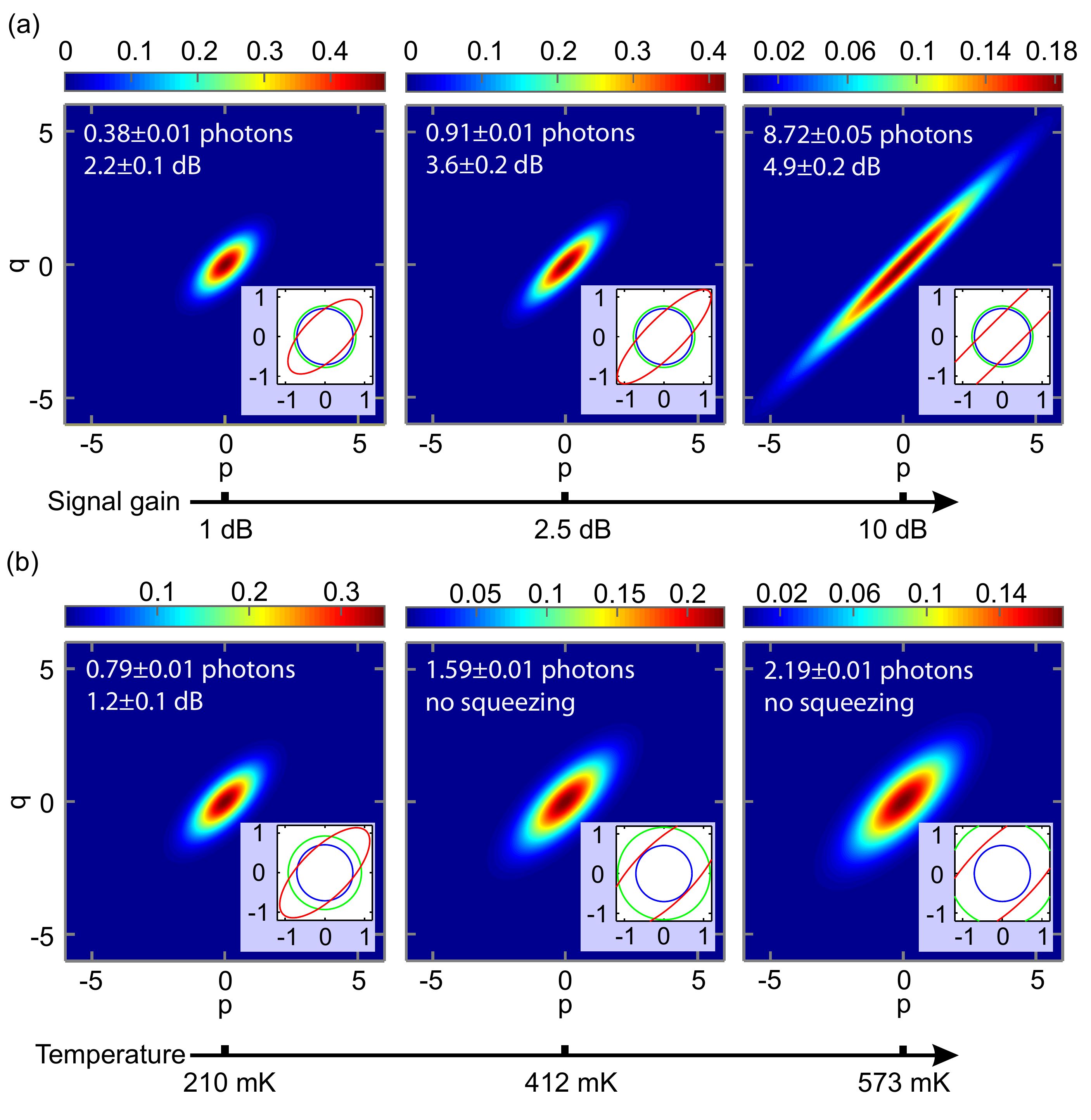}} 
\caption{
Reconstructed Wigner functions (color map) of squeezed vacuum and squeezed thermal states incident at the input port of the microwave beam splitter. $p$ and $q$ are dimensionless quadrature variables spanning the phase space. The insets show the $1/e$ contours of the ideal vacuum (blue), the experimental vacuum or thermal states (green), and the squeezed vacuum or squeezed thermal states (red). (a) Constant $30\,\deci\bel$-attenuator temperature of $50\,\milli\kelvin$. (b) Constant signal gain of $1\,\deci\bel$.  
}
\label{fig:Wigner}
\end{figure}
\begin{figure}[t!]
\begin{indented}
\item[]\includegraphics[width=0.8\columnwidth]{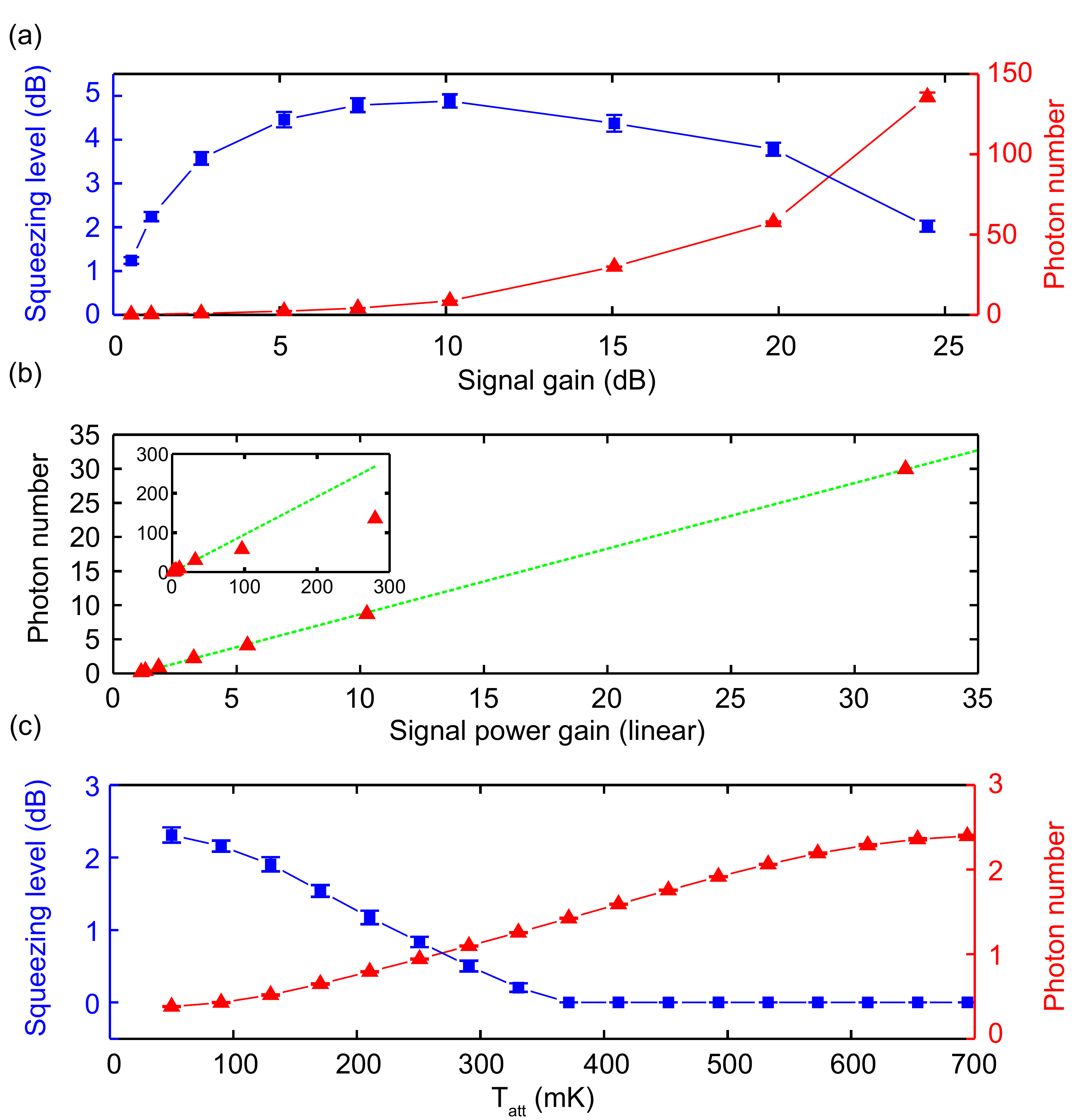}
\caption{(a)~Squeezing level ({\color{blue}\fullsquare}) and photon number ({\color{red} \large\fulltriangle}) plotted as a function of the signal gain when the $30\,\deci\bel$-attenuator is at $50\,\milli\kelvin$. The lines are guides to the eyes. 
(b)~Photon number as a function of signal power gain in linear units. Data: {\large\color{red}\fulltriangle}. Green dashed line: linear fit. Inset: Signal gain range equivalent to that shown in panel~(a). The two data points with the largest signal gain are excluded from the fit.
(c)~Squeezing level ({\color{blue}\fullsquare}) and photon number ({\color{red} \large\fulltriangle}) plotted as a function of the $30\,\deci\bel$-attenuator temperature for $1\,\deci\bel$ signal gain. The lines are guides to the eyes. All error bars are of statistical nature.
}
\end{indented}
\label{fig:SQdB}
\end{figure}

Selected Wigner function reconstructions of the state at the input of the beam splitter are plotted in figure~\ref{fig:Wigner}. When the $30\,\deci\bel$-attenuator is stabilized at $50\,\milli\kelvin$, vacuum fluctuations are present at the input of the JPA. As shown in figure~\ref{fig:Wigner}(a), the increase in pump power, which corresponds to an increase of signal gain, causes an increase of the squeezing level and an increase of the photon number (see also figure~\ref{fig:SQdB}(a)). We achieve a maximum squeezing level of $4.9\pm0.2\,\deci\bel$ below vacuum at $10\,\deci\bel$ signal gain. However, if we further increase the signal gain, the squeezing level decreases again. This behavior is expected~\cite{WallsMilburn} because the squeezing becomes suppressed when the JPA enters the bifurcation regime. In this regime, also the higher order cumulants do not vanish anymore. Indeed, we observe this effect in our data for signal gains larger than $10\,\deci\bel$. In addition, from the JPA input output relation~\cite{Scully} we identify the signal voltage gain in the non-degenerate mode in linear units as $G_{\rm signal,V}\,{=}\,\cosh\left(r\right)$. Applying this relation, we obtain from the photon number of a squeezed state, $n\,{=}\,\sinh^2\left(r\right)$, the expression $n\,{=}\,G_{\rm signal,P}-1$, where $G_{\rm signal, P}\,{=}\,G_{\rm signal, V}^2$ is the signal power gain in linear units. Therefore, we expect that the photon number increases linearly with $G_{\rm signal, p}$ in the non-degenerate mode with a slope of one. Figure~\ref{fig:SQdB}(b) confirms this behavior for small signal gains below the bifurcation regime. Next, we fix the signal gain at $1\,\deci\bel$. When the temperature of the $30\,\deci\bel$-attenuator is increased (see figure~\ref{fig:Wigner}(b) and figure~\ref{fig:SQdB}(c)), more and more thermal photons are incident at the input port of the JPA. Thus the squeezing level decreases and at some point the output state of the JPA is not squeezed below vacuum any more.

\section{Squeezed coherent states}
\label{section_squeezed_coherence}

\begin{figure}[ht]
\centering{\includegraphics[width=0.9\columnwidth]{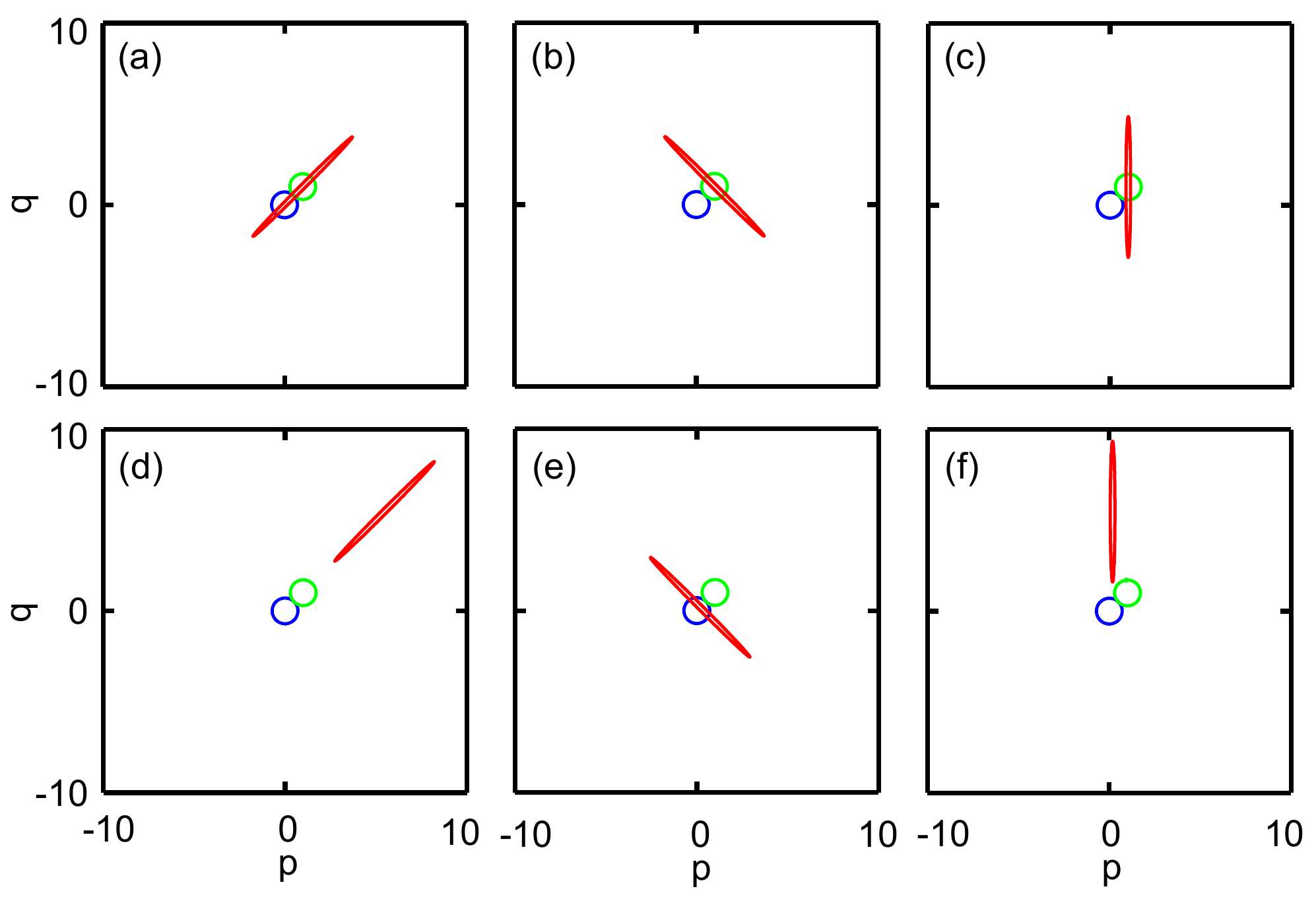}}
\caption{Sketch of $1/e$ contours of the ideal vacuum (blue), the coherent state (green), and the squeezed coherent state (red) with $r=1.7$, $\theta\,{=}\,45\degree$ and $\left | \alpha  \right |^2\,{=}\,2$. $p$ and $q$ are dimensionless quadrature variables spanning the phase space. (a)--(c)~Squeeze the vacuum state first, then displace. (d)--(f)~Displace the vacuum first, then squeeze. The anti-squeezed angle $\gamma$ is $45\degree$ in~(a) and (d),  $135\degree$ in~(b) and (e) and $90\degree$~in (c) and~(f).  
} 
\label{fig:squeezed coherent theory}
\end{figure}

\begin{figure}[ht]
\centering{\includegraphics[width=0.8\columnwidth]{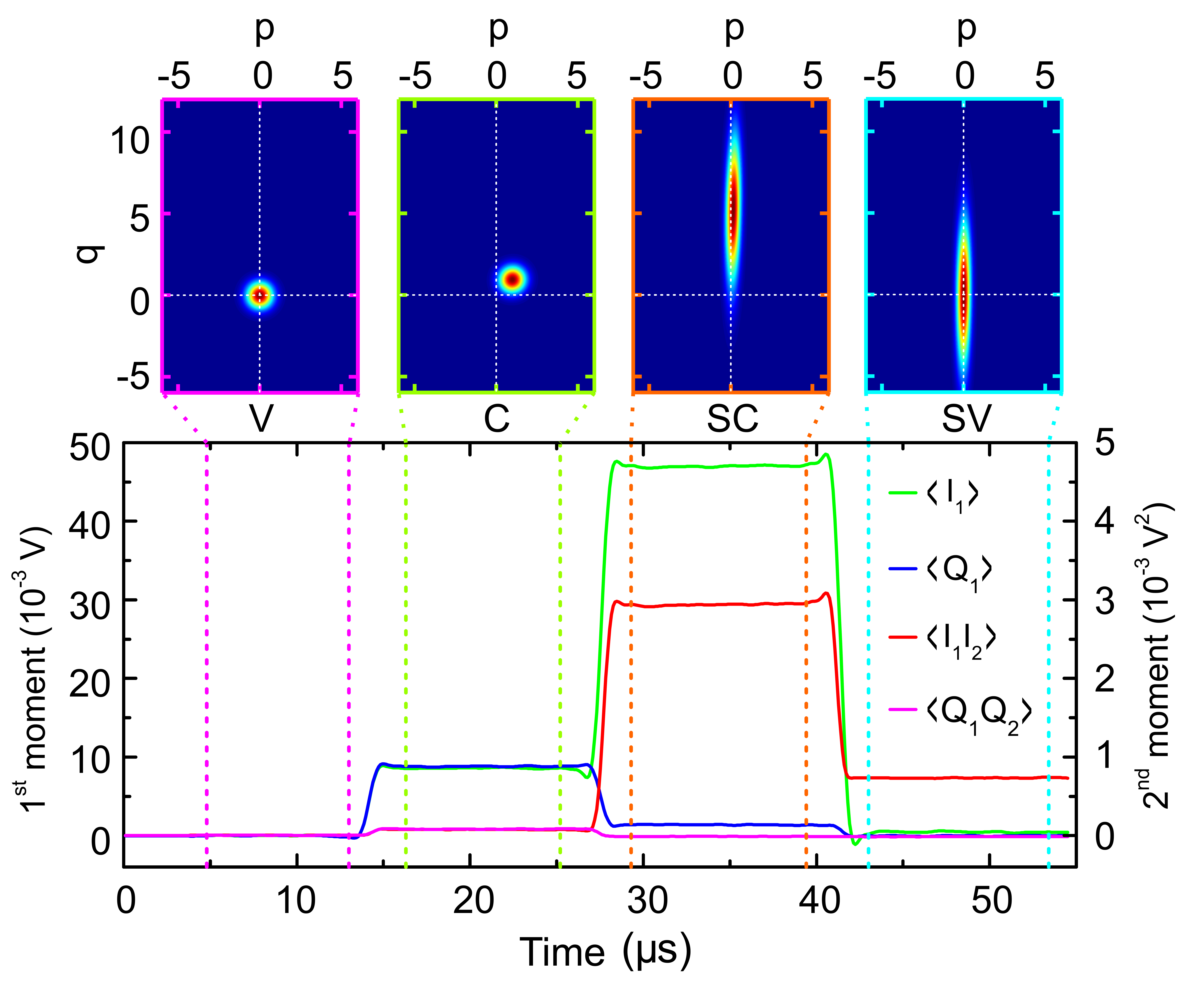}}
\caption{
Average over $5 {\times} 10^5$ traces of selected first and second moments from a squeezed coherent state measurement. The phase of the coherent state is $\Theta\,{=}\,45\degree$, and the anti-squeezed angle of the squeezed vacuum state is $\gamma\,{=}\,90\degree$. The four color maps above the time traces are the Wigner function reconstructions of the vacuum, coherent, squeezed coherent and squeezed vacuum states referred to the input of the beam splitter. $p$ and $q$ are dimensionless quadrature variables spanning the phase space.  
} 
\label{fig:TimeTrace}
\end{figure} 

\begin{figure}[ht]
\centering{\includegraphics[width=0.75\columnwidth]{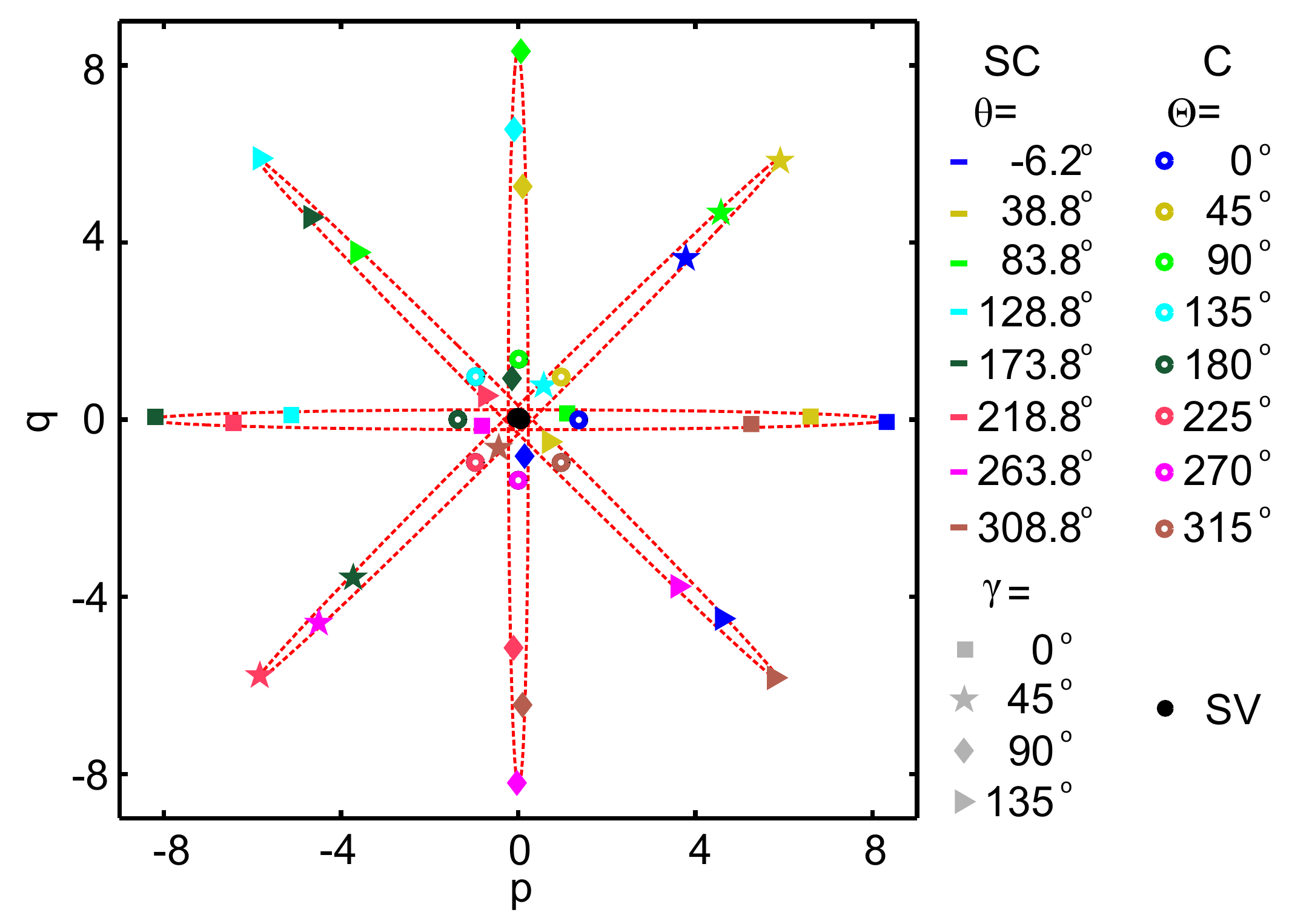}}
\caption{
Experimental displacement for squeezed vacuum, coherent and squeezed coherent states. The dashed red curves are fits of~(\ref{eq:displacement}) to the SC data. For each of the 32 $\Theta$- and $\gamma$-combinations, $2 {\times} 10^6$ traces are measured.
} 
\label{fig:Displacement}
\end{figure} 

\begin{figure}[ht]
\centering{\includegraphics[width=0.75\columnwidth]{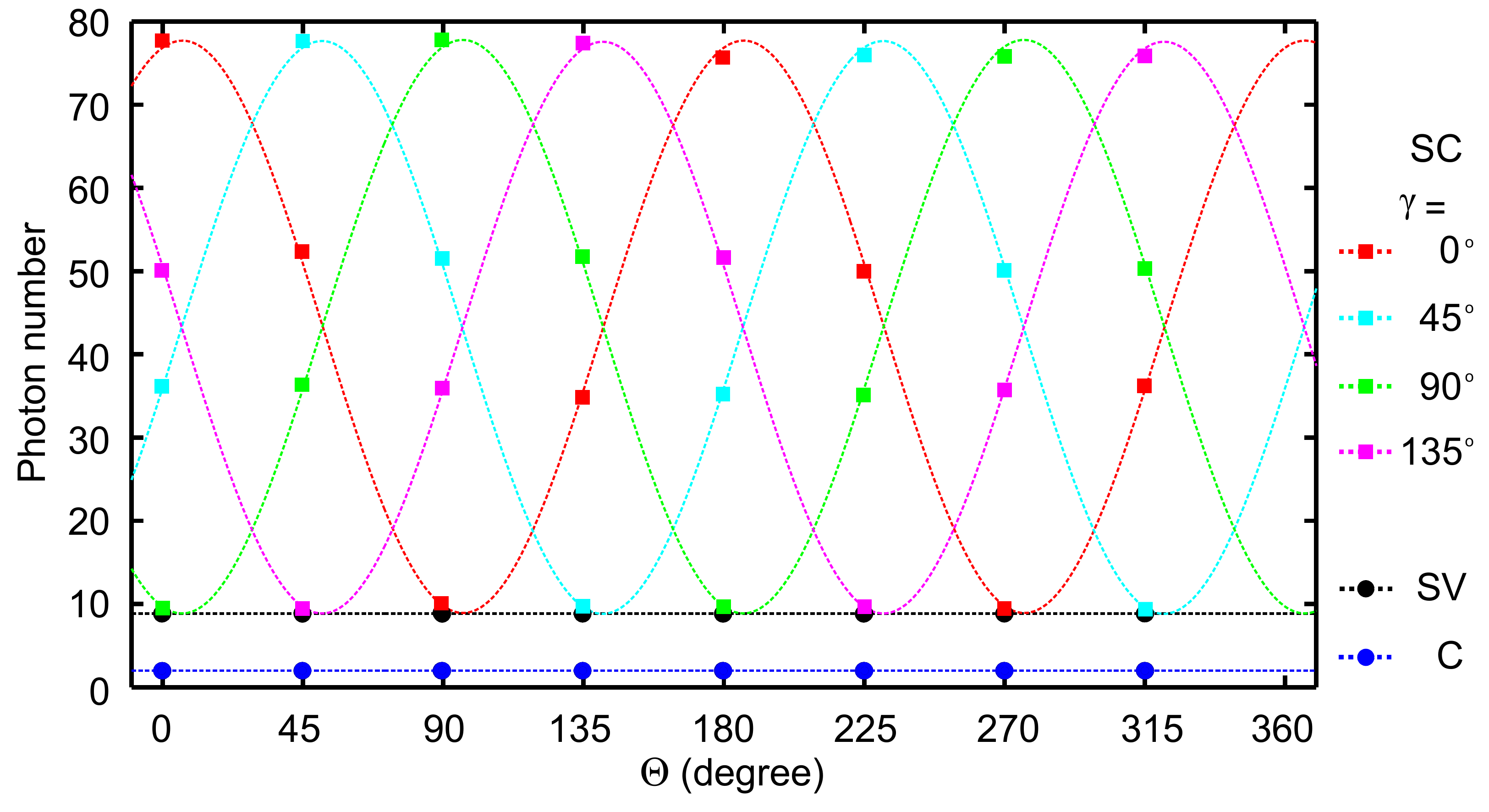}}
\caption{
Experimentally obtained photon numbers for coherent states, squeezed vacuum states and squeezed coherent states as a function of the coherent state phase $\Theta$. The statistical uncertainty is smaller than the symbol size. The dashed curves are fits using~(\ref{eq:Nphoton}). We note that the thermal contribution $\delta N\,{\approx}\,0.005$ is vanishingly small. Furthermore, although this fit is performed independently from the one in figure~\ref{fig:Displacement}, the obtained fit parameters are consistent. The data are from the same measurements as shown in figure \ref{fig:Displacement}.
} 
\label{fig:Nphoton}
\end{figure}

In addition to the squeezed vacuum and squeezed thermal states discussed in the previous section, we here analyze a more general squeezed state: the squeezed coherent state. There are two ways to obtain a squeezed coherent state. First, one can squeeze the vacuum state $\hat{S}\left(\xi\right)\left|0 \right\rangle$ and apply a displacement operation $\hat{D}\left(\alpha\right)\hat{S}\left(\xi\right)\left|0 \right\rangle$. Here, $\xi$ denotes the complex squeeze parameter $\xi\,{=}\,r \exp(\rmi\varphi )$ with squeeze factor $r$ and squeezing angle $\varphi$. Second, one can apply a displacement operator on the vacuum state $\hat{D}\left(\alpha\right)\left|0 \right\rangle$, and subsequently squeeze this displaced vacuum $\hat{S}\left(\xi\right)\hat{D}\left(\alpha\right)\left|0 \right\rangle$. We define the coherent phase $\theta$ as the angle between displacement direction and $p$-axis, and the anti-squeezed angle $\gamma\,{=}\,{-}\,\varphi/2$ as the angle between the anti-squeezed quadrature and the $p$-axis. We illustrate the difference of the two methods in figure~\ref{fig:squeezed coherent theory}. For the former method, the final displacement of the squeezed coherent state only depends on the displacement operation and is independent of the squeeze factor $r$ (figures~\ref{fig:squeezed coherent theory}(a)--(c)). However, the displacement of the squeezed coherent state obtained from the second method depends on both the displacement and squeezing operations. When the anti-squeezed quadrature is parallel to the displacement direction of a coherent state, the final displacement of the squeezed coherent state is maximum (figure~\ref{fig:squeezed coherent theory}(d)). Contrary, the final displacement reaches its minimal value when the anti-squeezed quadrature is perpendicular to the displacement direction (figure~\ref{fig:squeezed coherent theory}(e)). In this section, we present an analysis on squeezed coherent states generated with the second method and detected with the dual-path setup. To this end, we pump the JPA in the presence of a weak coherent signal. One measurement trace always records four regions: vacuum (V) region with both JPA pump and coherent signal off, coherent signal (C) region with JPA pump off and coherent signal on, squeezed coherent (SC) region with both JPA pump and coherent signal on, and squeezed vacuum (SV) region with JPA pump on and coherent signal off. 

Figure~\ref{fig:TimeTrace} shows typical averaged time traces of selected first and second moments from a single measurement, which corresponds to the situation in figure~\ref{fig:squeezed coherent theory}(f). Four Wigner function reconstructions are placed on top of the time trace. When both JPA pump and coherent signal are off, we can clearly identify the vacuum Wigner function with its maximum centered at the origin. Once we turn on the coherent signal, which is a displacement operation, the vacuum state is equally displaced along $p$- and $q$-quadratures and becomes a coherent state with phase $\Theta\,{=}\,45\degree$ referred to the beam splitter input. Next, keeping the coherent signal on, we turn on the JPA pump which gives a squeezing operation with the $p$-quadrature as the squeezed quadrature and the $q$-quadrature as the anti-squeezed quadrature ($\gamma\,{=}\,90\degree$). This results in a suppression of the displacement of the state along the $p$-quadrature and a corresponding amplification of the displacement along the $q$-quadrature. Also the $p$-quadrature variance is squeezed and the $q$-quadrature variance is amplified, turning the circular profile of the vacuum and coherent state Wigner function into an elliptic profile. Thus, the state becomes a squeezed coherent state with squeezing level $\cal S$$\,{=}\,4.3\,\deci\bel$. Finally, keeping the pump on, we turn off the coherent signal to generate a squeezed vacuum state with squeezing level $\cal S$$\,{=}\,4.7\,\deci\bel$. In this context, we would like to point to the following experimental aspect. Compared with a coherent state, whose phase $\Theta$ is referred to the beam splitter input, in the case of a squeezed coherent state, we use $\theta$ as the phase of the coherent state before applying the squeezing operator, and $\theta$ is referred to the JPA input. During the propagation of the coherent state from the input of the JPA to the input of the beam splitter, the phase of the coherent state evolves. We account for this effect by a constant phase difference $\Delta\Theta\,\equiv\,\Theta\,{-}\,\theta$. We note that the angle $\gamma$ always refers to the squeezed vacuum state recorded in each measurement trace. From theory~\cite{Scully}, we expect the displacement of a squeezed coherent state after the squeeze operation to depend on the angles as
\begin{equation}
 \left<\hat{a}\right>\,{=}\,\alpha \cosh r\,{-}\,\alpha^{*} e^{\rmi \varphi}\sinh r. 
 \label{eq:displacement}
\end{equation}
  Here, $\alpha\,{=}\,\left|\alpha\right |\exp\left[\rmi\pi\left(90\degree\,{-}\,\theta\right)/180\degree\right]$ is the complex amplitude of the coherent state before the squeeze operation and $\varphi\,{=}\,-2\gamma$ the angle of the complex squeeze parameter.

By fixing the anti-squeezing angle of squeezed vacuum states $\gamma$ at $0\degree$, $45\degree$, $90\degree$, and $135\degree$ and rotating the phase $\Theta$ of the coherent signal, we map out the dependence of the displacement of a squeezed coherent state on $\gamma$ and $\theta$. In figure~\ref{fig:Displacement}, we display the displacement, which is the center of the individual states in phase space given by their first moment $\left<\hat{a}\right>$ for various values of $\Theta$ and $\gamma$. The squeezed vacuum states are centered at the origin, and the coherent states are located on a circle around the origin. If we turn on the JPA pump and rotate the phase of a coherent signal, the squeezed coherent state moves mainly along the $\gamma$ direction. The displacement of the squeezed coherent states reaches its maximum when $\gamma\,{=}\,\theta\,{+}\,2n\times90\degree$, where $n\in \mathbb{Z}$. Geometrically, this means that the anti-squeezed direction is collinear to the displacement vector pointing from the origin to the center of the state. For our choice of $\gamma$, we obtain a characteristic star-shaped pattern.

Similar to the center of the Wigner function, which represents the displacement, the photon number of a squeezed coherent state varies when we rotate the phase of the coherent signal $\Theta$ while keeping the anti-squeezed angle $\gamma$ constant. Following~\cite{Scully}, we obtain:
\begin{eqnarray}
\fl\left \langle \hat{a}^{\dagger } \hat{a}\right \rangle=\left(\left | \alpha  \right |^{2}+\delta N\right)\left ( \cosh^{2}r+\sinh^{2}r \right )-\left ( \alpha ^{*}\right )^{2}\rme^{\rmi\varphi }\cosh r \sinh r\nonumber\\
- \alpha ^{2}\rme^{-\rmi\varphi }\cosh r \sinh r+\sinh^{2}r\label{eq:Nphoton},
\end{eqnarray}
where $\delta N$ describes the thermal photons present in the vacuum state at the JPA input and $\left | \alpha  \right |^2$ is the number of photons in the coherent state. As we see from figure~\ref{fig:Nphoton}, the photon number oscillates and reaches a maximum when $\gamma\,{=}\,\theta\,{+}\,2n\times90\degree$, $n\in \mathbb{Z}$. Thus, the photon number is maximal when the displacement of the SC state is maximal. We emphasize that the various states detected in our experiments are referred to the input of the beam splitter. To fit to theory, we need to shift the reference plane of the coherent state from the beam splitter input with phase $\Theta$ to the JPA input with phase $\theta$. Fitting the experimental data, we retrieve $\Delta\Theta\,{=}\,6.2\pm0.2\degree$ and a squeezing factor $r\,{=}\,1.8\pm0.1$. This $r$ value is quite close to the value $r_{\rm DP}\,{=}\,1.81\pm0.01$ measured for the individual states.

From theory, we expect that the only difference between a SC state and the corresponding SV state is a displacement in phase space without any rotation or deformation. Therefore, we analyze the statistics of the variances $\left(\Delta X_{\rm anti}\right)^2$ and $\left(\Delta X_{\rm sq}\right)^2$ of the anti-squeezed and squeezed quadratures and that of the angle $\gamma$ for the SC and SV states. As displayed in table~\ref{table:SCandSVcompare}, our data show that no significant rotations or deformations are present.

\begin{table}
\caption{\label{table:SCandSVcompare} Comparison between SC and SV states. The properties $B$ are obtained by pre-averaging over four values taken from moment reconstructions in the corresponding regions of $5{\times}10^5$ traces each. The root mean square is denoted as rms. The statistics is performed over the 32 $\Theta$- and $\gamma$-combinations shown in figures~\ref{fig:Displacement} and \ref{fig:Nphoton}.
}
\begin{indented}
\lineup
\item[]\begin{tabular}{lccc}
\br
$B$ & ${\rm rms}(B_{\rm SC}-B_{\rm SV})$ & ${\rm mean}(B_{\rm SC})\pm{\rm std}(B_{\rm SC})$ & ${\rm mean}(B_{\rm SV})\pm{\rm std}(B_{\rm SV})$\\
\mr
$\left(\Delta X_{\rm anti}\right)^2$ &0.43\0&$18.1\0\0\pm 0.3\0\0$	&$18.48\0 \pm 0.06\0$ \\ 
$\left(\Delta X_{\rm sq}\right)^2$  & 0.015 &$\00.189\pm 0.009$ & $\0 0.179\pm 0.005$\\
$\gamma$& 0.87\degree& -- & --\\
\br
\end{tabular}
\end{indented}
\end{table}

\section{JPA noise properties in degenerate mode}
\label{section_noise}

A JPA operated in the degenerate mode can not only generate vacuum squeezing, it can also be used as a low-noise phase-sensitive amplifier which, in principle, does not need to add any noise to the amplified quadrature\cite{Caves:1982}. With the dual-path setup, we therefore study the noise properties of our JPA in the degenerate mode. More precisely, we perform a temperature sweep of the 30\,\deci\bel-attenuator shown in figure~\ref{fig:DP setup}. The variance of the fluctuations at the frequency $f_0$ generated with this procedure is 
\begin{equation}
\left(\Delta X_{\rm therm}\right)^2=\frac{1}{4} \coth\left(\frac{h f_0}{2 k_{\rm B}T_{\rm att}}\right)
\label{eq:noise}
\end{equation}
where $\left(\Delta X_{\rm therm}\right)^2$ has the unit of photon number. At each temperature, the JPA pump is operated in the pulsed mode. Consequently, a single time trace in our measurement always contains a region corresponding to a non-squeezed thermal state and a region corresponding to a squeezed thermal state. For any quadrature, the variance $(\Delta X_{\rm out})^2$ at the output of the JPA is related to the variance $(\Delta X_{\rm in})^2$ at the input via the relation~\cite{Caves:1982}
\begin{equation}
\left(\Delta X_{\rm out}\right)^2=G_{\rm X}\left(\Delta X_{\rm in}\right)^2+\left(\Delta X_{\rm noise}\right)^2.
\label{eq:amplifier}
\end{equation}
Here, $G_{\rm X}$ is the gain for this quadrature and $\left(\Delta X_{\rm noise}\right)^2$ is the noise added by the amplifier referred to the output. In principle, we could determine the variance of the thermal state at the input of the JPA using the dual-path reconstructed signal moments at the input of the beam splitter taking into account the cable, circulator and JPA losses. However, the dual-path reconstruction detects a thermal population of 0.1 photons in the vacuum~\cite{PhysRevLett.109.250502} which would result in a significant underestimation of the JPA noise. For this reason, we calculate $\left(\Delta X_{\rm in}\right)^2$ based on~(\ref{eq:noise}), and model the cable loss between the $30\,\deci\bel$-attenuator output and the measurement circulator input with beam splitters, and account for the temperature gradients. In this way, we model an equivalent amplifier consisting of the measurement circulator, JPA, and cables to the beam splitter input. The noise contributions of all these components are represented by the noise temperature of the equivalent amplifier. Therefore, the latter is a pessimistic estimate for the noise properties of the JPA itself.   

\begin{figure}[ht]
\centering{\includegraphics[width=0.7\columnwidth]{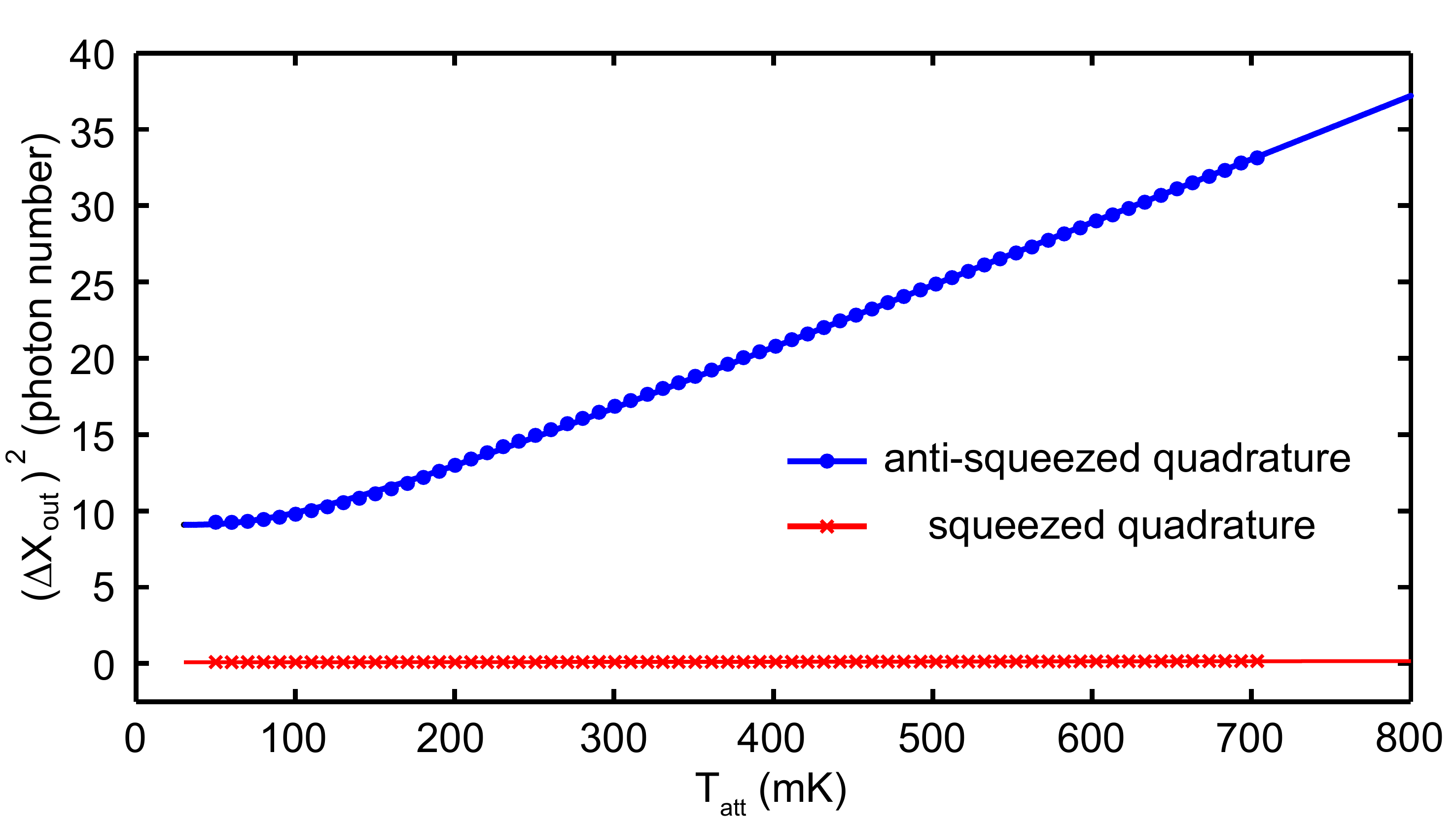}} 
\caption{
Anti-squeezed and squeezed quadrature variance as a function of the noise source temperature $T_{\rm att}$. Lines: Fits of of~(\ref{eq:amplifier}) to the data (symbols). The fit results are summarized in table~\ref{table:noise sum}. For each temperature point, $2 {\times} 10^6$ traces are measured.
}
\label{fig:noise DP}
\end{figure}

\begin{table}
\caption{\label{table:noise sum} Dual-path analysis results for JPA noise and gain along anti-squeezed and squeezed quadratures. The error bars describe the statistical error obtained from the fitting procedure.}
\begin{indented}
\lineup
\item[]\begin{tabular}{lcc}
\br
{}        						&\0$G_{\rm X}$ ($\deci\bel$)	& $\left(\Delta X_{\rm noise}\right)^2$\\ 
\mr
$X_{\rm sq}	$ 	& ${-}11.7\pm 0.3  $ 		& $0.06\pm 0.01$\\ 
$X_{\rm anti}$	& $\phantom{+}13.7\pm 0.1 $ 		& \0$0.14\pm 0.01^*$\\
\br
\end{tabular}
\item[] $\left(\Delta X_{\rm noise}\right)^2$ has the unit of photon number. \\
($^*$) Differently from the squeezed quadrature, where the added noise refers to the JPA output, this number refers to the JPA input for the anti-squeezed quadrature.
\end{indented}
\end{table}

In figure~\ref{fig:noise DP}, we show the anti-squeezed and squeezed quadrature variance as a function of the noise source temperature. Obviously, there is good agreement between theory and experiment. The corresponding numerical results are displayed in table~\ref{table:noise sum}. Most importantly, we observe a variance of $0.14\pm0.01$ photons (referred to the input) for the noise added by our JPA to the anti-squeezed quadrature. This value is clearly below the standard quantum limit of 0.25 photons for a single quadrature of a phase-insensitive amplifier. The relevant noise number for the use of the JPA as a squeezer is the noise it adds to the squeezed quadrature at the JPA output. We do not refer the added noise from the squeezed quadrature to the input of the equivalent JPA, because systematic uncertainties in the setup~\cite{PhysRevLett.109.250502} dominate the squeezed quadrature noise variance at the JPA output and would be amplified by $1/G_{\rm sq}$ when referring to the JPA input. However, these uncertainties are negligible for the anti-squeezed quadrature. Instead, following~\cite{Caves:1982} we calculate a lower bound for the squeezed quadrature noise variance at the JPA input from the experimentally more robust quadrature gains $G_{\rm sq}$ and $G_{\rm anti}$ of the squeezed and anti-squeezed quadrature and anti-squeezed quadrature noise variance. Using the values from table~\ref{table:noise sum}, we obtain the relation $\left(\Delta X_{\rm sq,in}\right)^2\,{\geq}\,\frac{1}{16}\left|1-\left(G_{\rm sq}G_{\rm anti}\right)^{-1/2}\right|^2/\left(\Delta X_{\rm anti, in}\right)^2\,{=}\,0.02$, where subscript ``in" indicates the JPA input.

\section{Conclusions and Outlook}
\label{section_conslusions}

In summary, we present a detailed analysis of the physics of squeezed microwave light generated with a flux-driven JPA. We first determine the operation point of the JPA and characterize its basic amplification properties, including non-degenerate gain, bandwidth, $1\,\deci\bel$-compression point and degenerate gain. We then use the JPA to squeeze vacuum fluctuations and find $4.9\pm0.2\,\deci\bel$ of squeezing at $10\,\deci\bel$ signal gain.
Furthermore, we investigate displacement and photon number of squeezed coherent microwave fields and find excellent agreement with theoretical expectations.
In the degenerate mode, we verify that our JPA, as a phase-sensitive device, adds less noise to the amplified quadrature than an ideal phase-insensitive amplifier. This property is of utmost importance for high efficiency detection, state tomography and quantum communication applications in the microwave domain.
 Furthermore, with the setup used in this work, i.e., squeezed states incident at a linear beam splitter, the generation of path entanglement between continuous-variable propagating quantum microwaves was demonstrated in~\cite{PhysRevLett.109.250502}. Considering recent work on the engineering of tunable beam splitter Hamiltonians~\cite{Peropadre2013,Reuther2010,Mariantoni2008}, our setup could be extended to the interesting case of dynamical switching between two separable single-mode squeezed states and a path-entangled two-mode squeezed state.
 
\ack
We thank C.~Probst, K.~Neumaier and K.~Uhlig for providing their expertise in cryogenic engineering. Furthermore, we acknowledge financial support from the Deutsche Forschungsgemeinschaft via the Sonderforschungsbereich~631, the German excellence initiative via the `Nanosystems Initiative Munich' (NIM), from the EU projects SOLID, CCQED, PROMISCE and SCALEQIT, from MEXT Kakenhi ``Quantum Cybernetics'', the JSPS through its FIRST Program, the Project for Developing Innovation Systems of MEXT, the NICT Commissioned Research, Basque Government IT472-10, Spanish MINECO FIS2012-36673-C03-02 and UPV/EHU UFI 11/55.

\section*{References}
\providecommand{\newblock}{}

\end{document}